%% file: main.tex
\documentclass[11pt]{article}
\usepackage{geometry}
\usepackage{makeidx}  
\usepackage{times}
\usepackage{framed}
\usepackage{newfloat}
\usepackage[caption=false]{subfig}
\usepackage{amsfonts}
\usepackage[many]{tcolorbox}
\usepackage{color}
\usepackage{cite}
\usepackage{enumerate}
\usepackage{xcolor}
\usepackage{pgfplots}
\usepackage{tikz}
\usepackage{url}
\usetikzlibrary{positioning}
\usetikzlibrary{arrows}
\usetikzlibrary{shadows.blur}
\usetikzlibrary{backgrounds}
\usepackage{mathtools}
\usepackage{amsmath}
\usepackage{algorithmic}

\usepackage{amsthm}
\usepackage[mathcal]{euscript}
\usepackage{float}
\usepackage[nolist]{acronym}
\usepackage{hyperref}
\usepackage[symbol]{footmisc}
\usetikzlibrary{arrows.meta}
\definecolor{linkgreen}{RGB}{0,130,0}
\hypersetup{
    colorlinks=true,       
    linkcolor=red,          
    citecolor=linkgreen,        
    urlcolor=linkgreen          
}
\usepackage{titlesec}
\usepackage{authblk}

\newcommand{\adv}{\mathcal{A}}

\newtheorem{definition}{Definition}
\newtheoremstyle{myremark} 
    {3pt}                    
    {1pt}                    
    {}                   
    {}                           
    {\bfseries}                   
    {.}                          
    {.5em}                       
    {} 
\theoremstyle{myremark}
\newtheorem{remark}{Remark}

\usepackage{todonotes}
\newcommand{\jci}[1]{ \todo[inline,color=magenta!20]{\textbf{JD:} #1 \color{black}}}

\newcommand{\smh}[1]{ \todo[inline,color=yellow]{\textbf{SMH:} #1 \color{black}}}

\newcommand{\ignore}[1]{\relax}
\title{\textbf{Beyond the Hype: On Using Blockchains in Trust Management for Authentication}\footnote{A version of this paper appeared at IEEE TrustCom-17 \url{http://ieeexplore.ieee.org/document/8029486/} \tiny{Copyright~\copyright~2017 IEEE}}}

\author{Nikolaos Alexopoulos}
\author{J\"org Daubert}
\author{Max M\"uhlh\"auser}
\author{Sheikh Mahbub Habib}

\affil{Telecooperation Lab\\
Technische Universit\"at Darmstadt\\
Hochschulstra{\ss}e 10, Darmstadt D-64289, Germany}

\date{}

\begin{document}

\renewcommand*{\thefootnote}{\fnsymbol{footnote}}
\maketitle
\renewcommand*{\thefootnote}{\arabic{footnote}}
\setcounter{footnote}{0}
\begin{center}
{\tt \small{\{alexopoulos, daubert, max, sheikh\}@tk.tu-darmstadt.de}}  
\end{center}
\vspace{5pt}
\begin{abstract}
	Trust Management (TM) systems for authentication are vital to the security
	of online interactions, which are ubiquitous in our everyday lives.
	Various systems, like the Web PKI (X.509) and PGP's Web of Trust are used
	to manage trust in this setting. In recent years, blockchain
	technology has been introduced as a panacea to our security problems,
	including that of authentication, without sufficient reasoning, as to
	its merits.

	In this work, we investigate the merits of using open distributed ledgers (ODLs),
	such as the one implemented by blockchain technology,
	for securing TM systems for authentication. We formally model such
	systems, and explore how blockchain can help mitigate attacks against
	them. After formal argumentation, we conclude that in the context of
	Trust Management for authentication, blockchain technology, and ODLs in general, can offer
	considerable advantages compared to previous approaches. Our analysis
	is, to the best of our knowledge, the first to formally model and argue about the security of TM
	systems for authentication, based on blockchain technology. To achieve this
	result, we first provide an abstract model for TM systems for authentication.
	Then, we show how this model can be conceptually encoded in a blockchain,
	by expressing it as a series of state transitions. As a next step, we
	examine five prevalent attacks on TM systems, and provide evidence that
	blockchain-based solutions can be beneficial to the security of such systems,
	by mitigating, or completely negating such attacks.
\end{abstract}
%
\input{intro}

\input{notation}

\input{model}

\input{attacks}

\input{challenges}

\section*{Acknowledgments}
This work has been co-funded by the DFG as part of project S1 within the CRC 1119 CROSSING and
as part of project D.4 within the RTG 2050 ``Privacy and Trust for Mobile Users''.

\input{acronyms}

\bibliographystyle{alpha}
\bibliography{sigproc}
\end{document}

%% file: intro.tex
\section{Introduction}\label{sec:intro}
We live in an increasingly interconnected world, where online interactions
are a common occurrence in the daily routine of most individuals.
Billions of interactions, possibly of a sensitive nature, take place
every day around the world. The security of these interactions is
essential for the orderly operation of organizations, as well as for the
well-being of people.

Authentication is a major enabler of security on the Internet.
In this context, authentication corresponds to a mechanism that verifies the
identities of interacting entities.
It is a requirement of almost all secure online interactions, 
ranging from email exchange to banking transactions. As an example, authentication
mechanisms as part of the TLS/SSL protocol \cite{dierks2008rfc} are used each time
a user browses a website through https \cite{rescorla2000http}. Through the use
of public key cryptography, authentication of
an entity equates to verifying the correctness of the binding between a public
key and an identity, which in our context also includes related attributes of the entity
(see section \ref{sec:id}). For the remainder of the paper this will
be called a \textit{public-key-to-id} binding.

Trust is a vital component of authentication systems. The electronic nature
of online interactions means, that physical identification using, e.g. state
issued id cards and passports, is not possible, or even appropriate. Furthermore, in many instances,
identification includes more than just name verification, e.g. in an online
marketplace scenario, the client would need to know if the interacting party
is actually the shop with the corresponding physical address that it claims,
possibly along with other relevant attributes, such as certifications of the shop's
quality.
Therefore, entities that have not had direct physical experience with the holders
of the public keys that they want to identify, need to rely on others to
authenticate the public-key-to-id bindings\footnote{We refer to systems that
	rely on trust for authenticating third parties as: \textit{``Trust
	Management systems for authentication''}}.
Entities trust that other entities are \textit{proficient} in the
identification process and \textit{honest} in the authentication process of
third parties. This trust is then disseminated in the network through chains
of trust expressed via cryptographic credentials, thus making 
possible the authentication of entities without physical interaction.
Two of the most widely used systems that achieve this goal are
the web X.509 PKI (\cite{chokhani2003internet,cooperrfc}) and PGP's Web of 
Trust \cite{zimmermann1995official}. The former follows the centralized
hierarchical approach, with designated root certification authorities 
(CAs) trusted a priori by browsers and operating systems. These
can in turn delegate this trust to other CAs and so on, to form chains.
The latter is based on a decentralized approach and allows users to 
authenticate other users by forming trust chains through their social
relations in the real world.

Real-world issues \cite{bright2011comodo, arthur2011diginotar, goodin2016firefox}
have highlighted the need for more secure and transparent authentication systems.
Recently, and in accordance with the need for decentralized and secure 
authentication, preliminary solutions using blockchain technology
have been proposed to solve some aspects of the 
problem (\cite{namecoin, fromknecht2014decentralized, slepak2013dnschain+, wilson2015pretty, ali2016blockstack, cryptoeprint:2016:1062})
with a varying degree of success. For example, Namecoin \cite{namecoin} offers a
decentralized namespace supported by a dedicated cryptocurrency. In
\cite{wilson2015pretty}, the authors propose storing PGP keys and signatures on the
bitcoin blockchain, while \cite{ali2016blockstack} introduces a distributed naming and
storage service that can be implemented on top of any suitable blockchain.
Blockchains and \acp{ODL} in general, as explained in Section \ref{sec:blockchain}, can
potentially offer 
security and transparency by design, to the problem, and therefore 
enhance the security of authentication infrastructures. However, the
solutions mentioned above, offer ad hoc advantages compared to the 
traditional schemes, that are in general not formally studied.

In this paper, we investigate whether or not, the hype \cite{dickson2017blockchain}
surrounding blockchain
technology, as a solution to security problems, has concrete backing. We limit
our investigation to \ac{TM} systems for authentication and we ask the following
question: \textit{Are there merits to using blockchain technology to enhance
	 the security of \ac{TM} systems for authentication? - and if so: which ones?}

To answer this question, we develop an abstract model for \ac{TM}
systems for authentication, based on graph theory. Our model's advantage
in comparison with previous work lies with its generic nature and simplicity.
It enables us to argue about \ac{TM} systems in a uniform way and express preperties
not specific to any one system.
We also provide a way to interpret the use of blockchain technology in this context,
by encoding our model to a state machine, corresponding to a secure blockchain.
Then, we present a set of five prevalent attacks against authentication systems, and
show that blockchain technology can in a great extent enhance the security of these
systems against those attacks. Therefore, as a result of formal reasoning, we 
conclude that the answer to the question posed above, is an affirmative one.
Our work is novel and significant in that regard, as this area has not been investigated 
in the past in a generic manner, and is an important step towards
understanding the importance of using this pervasive technology 
to its full potential.

\subsection{Our Contributions}
The main question we address with this paper is, whether or not blockchain technology
can enhance the security of \acl{TM} systems for authentication.
On the way to affirmatively answering the question above, we make a series of
contributions. Namely we present:
\begin{itemize}
	\item A graph theoretic model of trust networks for authentication and how to
		encode it in the blockchain.
	\item An analysis of selected attacks against trust networks and corresponging
		defenses enabled by blockchain technology.
	\item An insight into future directions for secure and decentralized
		\ac{TM}.
\end{itemize}
\subsection{Paper Organization}
We begin by going through notation and required preliminary knowledge in Section 
\ref{sec:preliminaries}. Then, we present our model in Section 
\ref{sec:model}, followed by an analysis of attacks and defenses
in Section \ref{sec:attacks}. Finally, we discuss our vision for future work in
this area in Section \ref{sec:challenges} and conclude in Section \ref{sec:concl}.

%% file: notation.tex
\section{Notation and preliminaries}\label{sec:preliminaries}
Before we introduce our model, we present the notation used throughout the paper, as well as some important background information.

\noindent\textbf{Notation}
A \textit{\ac{TM} system} for authentication is a system managing digital representations
of social trust to enable decisions regarding public-key-to-id bindings.
An \textit{entity} is a public key participating in the system. A \textit{chain}
is a path of directed edges of a (multi)graph. Moreover, $a \coloneqq b$ denotes that
$a$ is defined as $b$. The adversary is denoted with $\adv$.

\noindent\textbf{Public key cryptography}\label{sec:pkc}
Public key cryptography is the area of cryptography dealing with asymmetric
cryptographic systems. It is based on the notion that each entity has a pair of
cryptographic keys, a public one and a secret one. The public key of the entity can be used
to encrypt messages that only the possessor of the corresponding secret
key can decrypt. Furthermore, the possessor of a secret key can authenticate
messages by digitally signing them with it, and all other entities can verify
the signature with knowledge of the public key.

\noindent\textbf{Identity}\label{sec:id}
The Oxford dictionary defines \ac{id} as ``The characteristics determining who
or what a person or thing is''\footnote{https://en.oxforddictionaries.com/definition/identity}.
In the digital world, the identity associated with a public key can be perceived as a
set of attributes the holder of the corresponding private key has \cite{iso_identity}. In other words, who
or what the holder of the secret key is. For example, in the case of browsing the web,
identity corresponds to the domain name and the organization bound to a public key, whereas in the case of
email exchange, identity corresponds to the email address and person name bound to a public key. An
identity can be associated with multiple public keys, as a person may want to use
different keys for different purposes.

\noindent\textbf{Blockchain}\label{sec:blockchain}
Blockchain technology, as introduced with Bitcoin in \cite{nakamoto2008bitcoin},
offers an open, secure and distributed transaction ledger (\ac{ODL}).
As realizations of the technology focus on implementing
currency systems and are based on cryptographic primitives, they are known
as \textit{cryptocurrencies}. The basic idea behind the design is facilitating
decentralized consensus, that is, making it possible for a network of unknown
participants to jointly decide on a global view and ordering of transactions.
Transactions are grouped in blocks and in each round, a participant is elected
to propose a valid block. The election mechanism is in most cases
based on solving a computational puzzle (proof-of-work), but other approaches
like proof-of-stake \cite{buterin2013proof} have been proposed. Each participant
then inspects the proposed block and if it is valid, she includes it in her
blockchain.
We refer readers not familiar with blockchain technology to \cite{bonneau2015sok}
for an overview of bitcoin and cryptocurrencies.
Security of the blockchain is guaranteed under well defined assumptions,
relating to the ratio of honest 
and malicious users, as well as to the puzzle's hardness. Specifically, 
security in Bitcoin is guaranteed assuming an honest participating majority,
in addition with a puzzle hard enough so that the average time needed to solve it
is much less than the time needed for messages to propagate in the network.
The interested reader is referred to \cite{garay2015bitcoin} for further reading.
In this paper, we view the blockchain as a probabilistic state machine that converges
after a number of time steps, similar to how it is described in \cite{wood2014ethereum}
and \cite{saito2016s}. All entities view the blockchain as an append-only
history of state transitions.

%% file: model.tex
\section{A model of a TM system for authentication}\label{sec:model}
A general approach is necessary to address the 
\ignore{\jci{Here we need some of the content from Section 1. Why is a model important? What (level of modelling) did previous work use? What is missing, and why is the missing part so important?}
\smh{I think you should briefly talk about - why the TM system model for Authentication makes sense as a contribution - already in the Intro. The text in Intro. will motivate the reader to read this section. In this section, you present your model}}
common security issues of authentication \ac{TM} systems.
Previous attempts to model TM systems in other contexts use either graph theory \cite{josang2006simplification} or mathematical logic \cite{Habib2013TMCAIQ, martinelli2007relating}.
In this paper, we choose to use graph theory to model and reason about the security of such systems, as this representation is intuitive and can support arguments concerning the network connections of the nodes. Our representation is compatible with the seminal work of Maurer \cite{maurer1996modelling} and its extension \cite{marchesini2005modeling}.
We begin by defining the basic concepts of our model, then we consider
the capabilities of possible adversaries, and finally we give examples
by instantiating our model with OpenPGP and X.509 to show its generality.
\subsection{The Model}
We model a TM system for authentication, as a directed graph with vertices representing
entities that participate in the system, and edges representing trust
relations. Entities are perceived as public keys participating in the system.
A trust relation is the core basic concept of any TM system and
expresses the trust a trustor puts into a trustee.
\smallskip
\begin{definition}[Trust relation]
	A Trust relation ($TR$) is a sequence (tuple) $\langle A, B, c, v, \alpha, t \rangle$ where:
	\begin{itemize}
		\item $A$ is the trustor, i.e. the entity that expresses her trust 
			towards another entity.
		\item $B$ is the trustee, i.e. the entity that is the subject of the 
			trust expressed by $A$.
		\item $c$ is the context of the relation, i.e. an identifier of the 
			class the relation belongs to, along with relevant information.
		\item $v \in [0,1]$ is the actual trust value (assigned to $B$ by $A$. 
			In some cryptographic systems, like the X.509 PKI, it is 
			implicitly assumed 
			to be $1$ or $0$, i.e. total or no trust. Alternatively, it can
			take multiple
			discrete values, e.g. see Fig. \ref{openpgpmodel} for OpenPGP.
			In general, it can take arbitrary continuous or discrete 
			values, which in our model are normalized in the interval $[0,1]$.
		\item $\alpha$ is a set of cryptographic artifacts, most commonly 
			digital signatures, that are used to assure integrity and
		authenticity of the relation. 
		\item $t$ is a logical time component, marking the time the relation was
			last updated. We assume a partial ordering 
			of events in the distributed system \cite{lamport1978time}, 
			i.e. events at time $t+i,\: i>0$ are considered to have 
			happened after events at time $t$.
	\end{itemize}
\end{definition}
Regarding the context $c$ of the trust relation, for the purpose of authentication
two context classes are generally relevant. First, the context class of \textit{validity}.
In this context class, $c$ includes an identity, i.e. a set of attributes, and
the trust relation expresses the belief that $A$ has about the binding of the
identity in $c$, with $B$. The second relevant context class, the \textit{authenticator
trust} context class, expresses the trustworthiness of $B$ as an authenticator
of unknown entities. It is possible that $A$ trusts $B$ in the context of validity,
i.e. she knows the identity of $B$, however she does not trust $B$ to authenticate
other entities. Hence, differentiating among different trust contexts is necessary.
A trust relation can also express negative trust information, like the revocation
of a key, when e.g. the secret key is lost or the identity of an entity changes.
Revocation information is represented by setting $v=0$. Complete uncertainty can be
modelled by setting $v=0.5$.

To express the whole ecosystem of trust relations, we define the \ac{TM} network, which
includes all trust relations between participating entities.
\smallskip
\begin{definition}[TM network]
	A TM network or trust graph is a directed multigraph $G = (V, E)$ where:
	\begin{itemize}
		\item Each $\text{v} \in V$ is an entity, e.g. certification authority, 
			physical person etc.
		\item Each $e \in E$ is labeled with a trust relation.
	\end{itemize}
\end{definition}
\smallskip
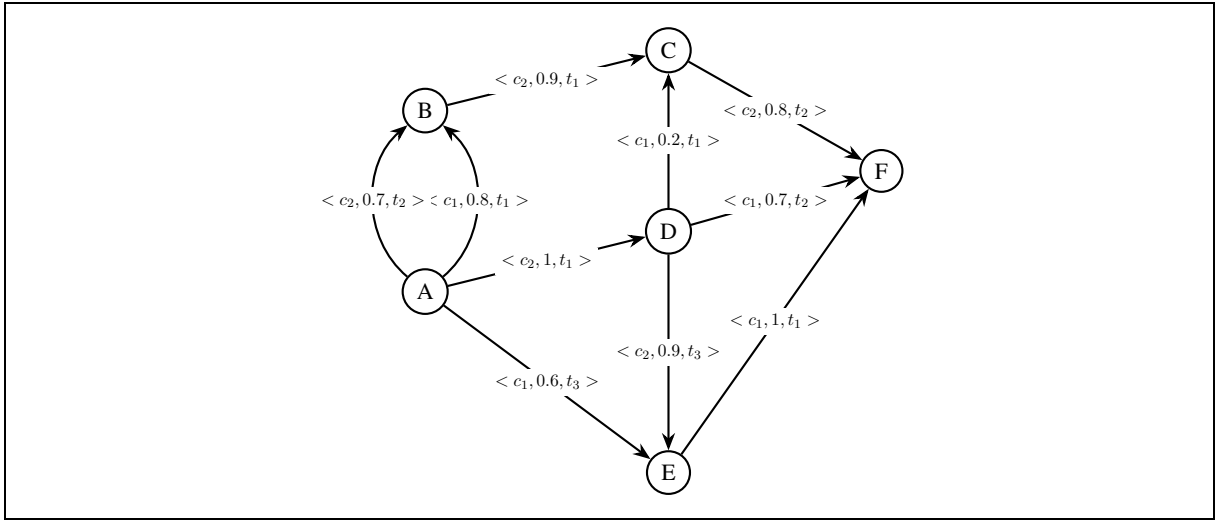
\begin{figure}
\caption{An example TM network. We omit the two first members of the tuple,
as the information is conveyed by the graphical representation. Also,
$\alpha$ is omitted for brevity. The context $c$, is depicted
with its class as subscript. We can see a variety of trust
relations of different contexts and timestamps, forming a (multi)graph.}
\label{fig:TMnet}
\smallskip
\begin{framed}
\centering
\begin{tikzpicture}[scale=0.8]
\begin{scope}[every node/.style={circle,thick,draw,scale=0.8}]
	\node (A) at (-3,0) {A};
    	\node (B) at (-3,3) {B};
    	\node (C) at (1,4) {C};
    	\node (D) at (1,1) {D};
    	\node (E) at (1,-3) {E};
    	\node (F) at (4.5,2) {F} ;
\end{scope}
\begin{scope}[>={Stealth[black]},
              every node/.style={fill=white,rectangle,scale=0.6},
              every edge/.style={draw=black,thick}]
	      \path [->] (A) edge[bend right=50] node {$<c_1,0.8,t_1>$} (B);
    \path [->] (A) edge[bend left=50] node {$<c_2,0.7,t_2>$} (B);
    \path [->] (B) edge node {$<c_2,0.9,t_1>$} (C);
    \path [->] (A) edge node {$<c_2,1,t_1>$} (D);
    \path [->] (D) edge node {$<c_1,0.2,t_1>$} (C);
    \path [->] (A) edge node {$<c_1,0.6,t_3>$} (E);
    \path [->] (D) edge node {$<c_2,0.9,t_3>$} (E);
    \path [->] (D) edge node {$<c_1,0.7,t_2>$} (F);
    \path [->] (C) edge node {$<c_2,0.8,t_2>$} (F);
    \path [->] (E) edge node {$<c_1,1,t_1>$} (F); 
\end{scope}
\end{tikzpicture}
\end{framed}
\end{figure}
An example of a \ac{TM} network is given in Fig. \ref{fig:TMnet}.
Finally, the goal of a TM network is to enable the assessment of trust
in different contexts, by taking into account the propagation of trust along
edges.
\smallskip
\begin{definition}[Trust assessment]
	A trust assessment $T_{A \rightarrow B}^c$ is the
	result of the calculation of trust $A$ puts into $B$ in a given
	context $c$, defined as:
	\begin{equation*}
		T_{A \rightarrow B}^c \coloneqq \mathcal{P} (c, H),\quad \mathcal{P} : c \times H \subseteq G \rightarrow [0, 1]
	\end{equation*}
	where $\mathcal{P}$ is a program that takes as input a trust network in the form 
	of a network subgraph $H$, which represents the view of the graph by $A$,
	and outputs a trust value in a given context.
	This trust value can again take binary, 
	discrete or continuous values and is normalized in the interval $[0, 1]$
	. The subgraph $H$ is the \textit{trust view} 
	of the trust assessment and includes the vertices and edges used for the computation 
	by the program $\mathcal{P}$. The program $\mathcal{P}$ expresses the way trust
	propagates in the network, e.g. via trust chains.
\end{definition}
\subsection{The Adversary}
In this paper, we consider an adversary $\adv$ that attacks the \ac{TM} system
in order to impersonate an honest entity, i.e. claim an identity that is not rightfully
his. In computer security terminology, this is a \ac{MITM} attack. Depending on the
attack, the adversary might also want to remain undetected.
We assume a powerful adversary that 
can arbitrarily add and remove labeled edges and vertices from the network 
graph $G$. In literature, this adversary is known as an \textit{arbitrary adversary} 
\cite{mitzenmacher2016models} and is equivalent to an adversary that controls a 
subset of the entities of the system, along with a subset of the communication 
channels. Finally, we assume that
there exist cryptographic primitives, namely digital 
signatures and encryption schemes, that are secure against $\adv$, i.e. the
adversary is computationally bounded as per standard security literature.  
\subsection{Trust assessment security}
As TM systems are generally used to assist in decision making, their security 
could be defined as, whether or not the decisions made using them 
are the correct ones. For example, if access to a resource is only granted to
an authorized party, or a given public key actually belongs to a physical 
entity. The latter definition is the one we adopt in this
paper. However, these decisions depend not only on the TM network 
infrastructure, but also on other factors, like the program $\mathcal{P}$ 
used to derive the trust assessment, as well as the decision strategy 
of the user. Therefore, the security of a \ac{TM} system 
is conceptually decoupled from the decision made, and resolves 
around computing a correct trust assessment, even when some entities 
and communication links are controlled by an adversary $\mathcal{A}$. 
\subsection{Examples}
We showcase the generality of our model by fitting it to two representative
systems following different approaches.
\subsubsection{OpenPGP Web of Trust}
The Open Pretty Good Privacy protocol (OpenPGP) is an open source version
of the initial PGP protocol and the de facto standard for decentralized
exchange of encrypted messages. It employs an anarchic Web of Trust (WoT)
to (potentially mutually) authenticate parties.
Trust information is stored in specified keyservers that are tasked with
making it available upon request. More details about the state and the security of the OpenPGP WoT can be found
in \cite{ulrich2011investigating} and \cite{barenghi2015challenging}. The realization
of our model for the OpenPGP WoT can be seen in Fig. \ref{openpgpmodel}.
\begin{figure}[h]
\caption{OpenPGP Web of Trust instantiation}\label{openpgpmodel}
\begin{framed}
\begin{enumerate}[i.]
	\item Trust relations are signed certifications where:
	\begin{itemize}
		\item $A$ is the trustor, i.e. the entity that expresses trust
		\item $B$ is the trustee, i.e. the subject, toward which trust is expressed.
		\item $c \in \{\text{validity, trust}\}$. The system employs two 
			contexts, one for expressing trust in the validity of a
			public-key-to-id binding and another expressing the level of trust
			in an entity as an authenticator.
		\item $v$ can take different values according to the context. For the 
			validity context, it takes values
			$v \in \{\text{full, marginal, untrusted, unknown}\}$ and for the trust 
			context it takes values $v \in \{\text{ultimate, full, marginal, untrusted,\\ 
			undefined}\}$.
		\item $\alpha$ is a digital signature verifying $A$ is the owner
			of the relation.
	 	\item $t$ is the system time of the last update of the relation
	\end{itemize}
	 \item The trust assessment concerns the validity of a given \textit{public-key-to-id} 
		 binding. More specifically, program $\mathcal{P}$ outputs the level of trust 
		 put on the validity of a binding, by forming certification chains. By
		 default, PGP requires one fully (or ultimately) trusted signature or
		 two marginally trusted ones to establish a key as valid. More details
		 on the system are available in \cite{callas2007rfc}.
\end{enumerate}
\end{framed}
\end{figure}
\smallskip
\subsubsection{X.509 PKI}
The X.509 standard PKI is a representative of the centralized model for
authentication and embraces a hierarchical trust structure. It is the
standard method for authenticating web domains, and therefore the most
widely used authentication system today. Several ``root'' certification
authorities are pre-trusted by default by browser vendors, e.g. Mozilla
products ship with $\approx170$ root CAs \cite{mozillaca}.
These root CAs can in turn delegate this trust to any other CA, and any CA
can authenticate any given identity (usually domain name). The realization
of our model for the X.509 PKI can be seen in Fig. \ref{x509model}
\begin{figure}[h]
\caption{X.509 PKI instantiation}\label{x509model}
\begin{framed}
\begin{enumerate}[i.]
	\item Trust relations are signed certifications where:
	\begin{itemize}
		\item $A$ is the trustor, i.e. a user that trusts a
			CA, or a CA.
		\item $B$ is the trustee, i.e. the subject toward which trust is expressed.
		\item $c \in \{\text{certify, authenticate}\}$. The system employs two 
			contexts. One for expressing the delegation of trust from a
			CA of a higher tier to one of a lower tier (certify) and
			one for expressing the trust a CA puts on a public-key-to-id
			binding of an entity-domain (authenticate).
		\item $v$ is set to $1$, i.e. total trust, for all trust relations expect revocation information, in which case\\ $v=0$. 
		\item $\alpha$ is a digital signature verifying $A$ is the owner
			of the relation.
	 	\item $t$ is the system time of the last update of the relation
	\end{itemize}
	 \item The trust assessment concerns the validity of a given \textit{public-key-to-id} 
		 binding. Program $\mathcal{P}$ works as follows: if there exists any chain of trust from an entity to another
		 entity, where all but the last relation are of a certification context,
		 i.e. $c=\text{certify}$, and the last relation of the chain
		 is of authentication context, i.e. $c=\text{authenticate}$, then the
		 trust put into the public-key-to-id binding is $1$.
\end{enumerate}
\end{framed}
\end{figure}
\smallskip
\begin{remark}
	The goal of this section is not to exhaustively specify existing
	authentication systems in our model, but to showcase the generality
	of our abstraction.
\end{remark}
\subsection{Modelling the use of blockchain}\label{sec:blockmodel}
\begin{figure}
	\caption{Example of graph transitions on the blockchain. The initial state is the empty
	graph $G_0 = (\emptyset, \emptyset)$. The state machine has converged up to a
point $t$ in the future. A fork starts at block $t+1$.}\label{fig:blockchain}
\smallskip
\begin{framed}
\centering
\begin{tikzpicture}[scale=0.8]
\begin{scope}[every node/.style={rectangle,thick,draw,scale=0.8}]
	\node (A) at (0,0) {$\phantom{a}G_0\phantom{a}$};
	\node (B) at (2,0) {$\phantom{a}G_1\phantom{a}$};
	\node[draw=none] (C) at (4,0) {$\cdots$};
	\node (D) at (6,0) {$\phantom{a}G_t\phantom{a}$};
	\node (E) at (8,0.5) {$G'_{t+1}$};
	\node (F) at (8,-0.5) {$G''_{t+1}$};
\end{scope}

\begin{scope}[>={Stealth[black]},
              every node/.style={fill=white,circle,scale=0.8},
              every edge/.style={draw=black,thick}]
    \path [->] (A) edge (B);
    \path [->] (B) edge (C);
    \path [->] (C) edge (D);
    \path[dashed] [->] (D) edge (E);
    \path[dashed] [->] (D) edge (F);
\end{scope}
\end{tikzpicture}
\end{framed}
\end{figure}
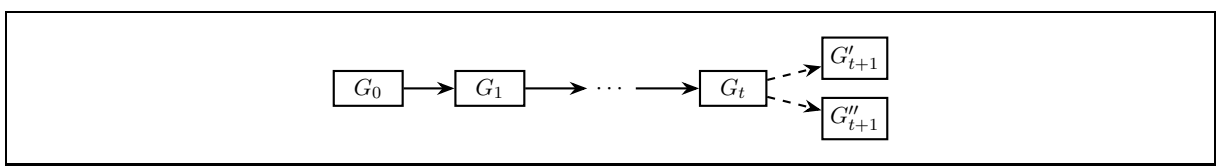

After describing our graph model for authentication \ac{TM} systems,
we proceed to provide a blockchain encoding of it.
As mentioned in Section \ref{sec:blockchain}, we model the blockchain
as a tamper proof, distributed state machine. Trust transactions either add,
remove or modify an edge of graph $G$. Vertex existence is
implied by the existence of edges. Each block committed to the
blockchain represents a snapshot of the graph at a given time $t$,
as it can be seen in Fig. \ref{fig:blockchain}.
The snapshot is considered stable, i.e. tamper proof, after an adequate
amount of time steps, e.g. in Bitcoin a transaction is assumed irreversible
after 6 confirmations as a rule of thumb \cite{bitcoinconfirm}. In our
reasoning about the use of blockchain, we assume entities accept only
tamper proof graph snapshots, i.e. they take into account state transitions that are
adequately deep in the chain, and therefore can be reversed with negligible
probability.

To sum up, in this section, we introduced our working model of \ac{TM} systems for
authentication and how it can be conceptually realized by a blockchain.
This lays the ground for the attacks and defences that follow in Section
\ref{sec:attacks}.

%% file: attacks.tex
\section{Attacks and defenses}\label{sec:attacks}
Attacks described in this section, and generally attacks against authentication systems, are characterized as \textit{impersonation attacks}.
The goal of the adversary is to impersonate another entity, usually
performed via a MITM attack.
In this section, we present common and realistic attacks
against authentication systems, and how blockchain technology enables
defending against them. For an insight into other methods to mitigate these attacks
we refer the readers to remark \ref{rem:other}. We employ example graphs to better illustrate the attacks.
The graphs follows the notation of Section \ref{sec:model}, with the addition
of dashed lines which represent trust relations that are part of a chain
affected by the attack. Context notation is suppressed for simplicity.
Colored vertices represent entities controlled by $\adv$.
\subsection{Stealthy targeted attack}
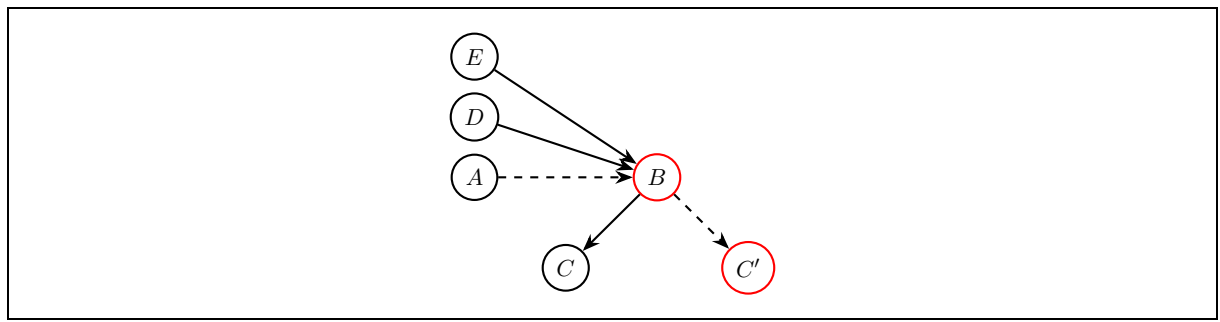
\begin{figure}[h]
	\caption{Stealthy targeted attack example}\label{fig:a1}
	\smallskip
	\begin{framed}
	\centering
	\begin{tikzpicture}[scale=0.8]
	\begin{scope}[every node/.style={circle,thick,draw,scale=0.8}]
		\node (A) at (0,0) {$A$};
		\node[draw=red] (B) at (3,0) {$B$};
    		\node (C) at (1.5,-1.5) {$C$};
		\node (D)[draw=red] at (4.5,-1.5) {$C'$};
    		\node (E) at (0,1) {$D$};
    		\node (F) at (0,2) {$E$};
	\end{scope}

	\begin{scope}[>={Stealth[black]},
		every node/.style={fill=white,circle,scale=0.8},
	every edge/.style={draw=black,thick}]
	      	\path [->] (A) edge[dashed] (B);
	      	\path [->] (B) edge (C);
	      	\path [->] (B) edge[dashed] (D);
	      	\path [->] (E) edge (B);
	      	\path [->] (F) edge (B);
	\end{scope}
	\end{tikzpicture}
	\end{framed}
\end{figure}

\noindent\underline{The attack:}
In this scenario, an adversarial entity forwards different trust relation information about
a given subject, to the attack target, compared to the rest of the network.
This way, $\adv$ manipulates the target's trust assessment.
An attack of this kind aims at mounting a MITM against a specific
user while avoiding detection. The adversary
controls a sufficient number of entities, that are directly or 
indirectly trusted by the target. An example of this class of attacks would
be a malicious CA in a X.509 system, that wants to gain access to
private information of a specific user.
This attack is similar in nature to what is known as a \textit{discrimination}
\cite{josang2009challenges, wang2014towards} or \textit{conflicting behavior} \cite{sun2006trust}
against trust evaluation
systems. By providing malicious trust information only to a specific target,
the adversary mounts a \ac{MITM} attack against the target, that is difficult to detect, hence stealthy.
An example can be seen in Fig. \ref{fig:a1}, where $\adv$ controls $B$ and
mounts a \ac{MITM} against $A$, by maliciously authenticating $C'$ as having
the identity that is associated with $C$.
The rest of the \ac{TM} network, i.e. $D$ and $E$, continue to receive correct information from
$B$ and therefore cannot detect the attack.
\smallskip\\\noindent\underline{The defense:}
The success of attacks of this kind is based on the inability of a
large portion of the network to detect the attack. Under the assumptions of our
blockchain model for trust management \ref{sec:blockmodel}, this attack is
no longer possible. This is the case because having conflicting views
of the trust relations at a given time $t$ would mean having conflicting
views of the trust graph $G$, which in turn would mean that a blockchain
fork is present starting at a time $t'\leq t$. According to our assumptions
about the security of the blockchain, this event is improbable, as entities are
assumed to take into account blocks adequately deep in the chain. Therefore,
the attack cannot take place with non negligible probability.
\subsection{Double registration attack}
\begin{figure}[h]
	\caption{Double registration attack example}\label{fig:a2}
	\smallskip
	\begin{framed}
	\centering
	\begin{tikzpicture}[scale=0.8]
	\begin{scope}[every node/.style={circle,thick,draw,scale=0.8}]
		\node (A) at (0,0) {$A$};
		\node[draw=red] (B) at (3,0) {$B$};
		\node (C')[draw=red] at (1.5,-1.5) {$C'$};
    		\node (D) at (0,1) {$D$};
    		\node (E) at (0,2) {$E$};
    		\node (F) at (7,2) {$F$};
    		\node (G) at (7,1) {$G$};
    		\node (H) at (7,0) {$H$};
    		\node (I) at (4,0) {$I$};
    		\node (C) at (5.5,-1.5) {$C$};
	\end{scope}

	\begin{scope}[>={Stealth[black]},
		every node/.style={fill=white,circle,scale=0.8},
	every edge/.style={draw=black,thick}]
	\path [->] (A) edge[dashed] (B);
	      	\path [->] (B) edge[dashed] (C');
		\path [->] (D) edge[dashed] (B);
		\path [->] (E) edge[dashed] (B);
	      	\path [->] (F) edge (I);
	      	\path [->] (G) edge (I);
	      	\path [->] (H) edge (I);
	      	\path [->] (I) edge (C);
	\end{scope}
	\end{tikzpicture}
	\end{framed}
\end{figure}
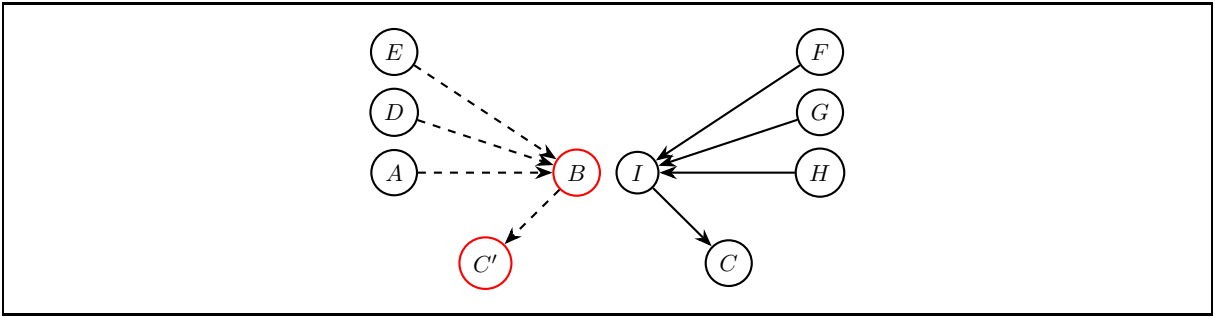

\noindent\underline{The attack:}
In this scenario, the adversary wants to mount an attack against a set of targets
by stealing the identity of an entity already participating in the system.
An example would be, an entity with considerable capabilities, that succeeds in
convincing a set of honest entities to authenticate the former as
having the identity of a different user, already participating in the system.
This impersonation attack could be achieved by issuing fake government
credentials or tricking the authenticators into believing e.g. that
a person represents an organization when this is not the case. An
example of the attack can be seen in the \ac{TM} network of Fig. \ref{fig:a2}, where entity
$B$ authenticates $C'$ as having an identical identity to $C$. The
partition of the trust network, makes the attack undetectable in the long
run and entities $A,D,E$ are affected, leading them to make wrong trust assessments. We note that binding two public keys
with the same identity is not an attack by itself. To the contrary, it is
a well established practice for individuals to have more than one public keys
for different purposes, e.g. one key for signature verification and another one
for encryption. However, the fact that knowledge of the different keys is not
shared with the entire network is what makes this scenario an attack.
\smallskip\\\noindent\underline{The defense:}
The success of this attack is based on the limited scope of view of
the participating entities. Specifically, there exist two trust relations
binding different public keys to the same identity and no honest entity has knowledge
of both. Under the assumptions of our blockchain
model, all participants have global view of the trust relations, i.e.
they know the whole trust graph. Thus, two trust relations with the properties
described above will be detected by every participant of the system and therefore
this attack is not possible. Detecting the attack naturally negates it, as
the target should not consider the information provided by the attacker as
valid.
%
%
\ignore{
\subsection{Nothing to lose attack}
\noindent\underline{The attack:}
In this scenario, the adversary-controlled entities behave honestly for
a period of time so that they build solid trust relations and then
decide to mount \ac{MITM} attacks against a large set of targets without
taking precautions against the detection of the attack. This scenario
could be realized by the compromise of a CA's secret key. This attack
can be related to the \textit{on-off} attack \cite{sun2006trust}
or the \textit{reputation lag exploitation} attack \cite{josang2009challenges}
of the generic trust management literature.}
%
%
\subsection{Stale information attack}
\begin{figure}[h]
	\caption{Stale information attack example}\label{fig:a3}
	\smallskip
	\begin{framed}
	\centering
	\begin{tikzpicture}[scale=0.8]
	\begin{scope}[every node/.style={circle,thick,draw,scale=0.8}]
		\node (A) at (0,0) {$A$};
		\node[draw=red] (B) at (2,0) {$B$};
    		\node (C) at (4,0) {$C$};
    		\node (E) at (6,0) {$E$};
	\end{scope}

	\begin{scope}[>={Stealth[black]},
		every node/.style={fill=white,circle,scale=0.8},
	every edge/.style={draw=black,thick}]
	      	\path [->] (A) edge[dashed] (B);
		\path [->] (B) edge[dashed] (C);
		\path [->] (C) edge[bend right, dashed] (E);
		\path [->] (C) edge[bend left] (E);
	\end{scope}
	\end{tikzpicture}
	\end{framed}
\end{figure}
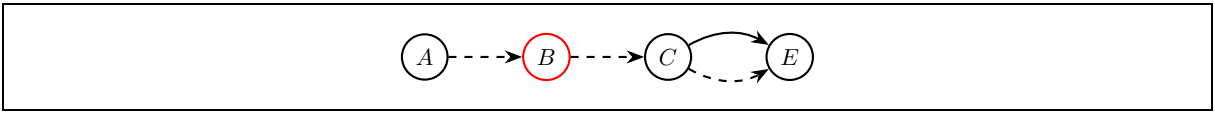

\noindent\underline{The attack:}
In this scenario, an entity makes a trust assessment based on
information that is not up to date. The most common occurrence
of this attack is when an entity makes a decision regarding the
authentication of another entity, without taking into account
relevant revocation information. In the example of Fig. \ref{fig:a3},
$B$ forwards to $A$ a trust relation of $C$ that is not up-to-date.
Specifically, there exist trust relations $TR_1 = \langle B,C,c_1,v_1,\alpha_1,t_1 \rangle$
and $TR_2 = \langle B,C,c_2,v_2,\alpha_2,t_2 \rangle$ with $t_1<t_2$, so that
$A$ is aware of $TR_1$ but not $TR_2$. In the case that $TR_2$ expresses
negative information about the validity of the key binding of $E$, then
this attack can lead to a \ac{MITM} against $A$. Imagine that $C$ is a
CA that revokes its authentication of $E$ due to, e.g. the theft of $E$'s
secret key. Then, the adversary, by withholding this information from $A$
leaves him vulnerable to an attack by the thief.
\smallskip\\\noindent\underline{The defense:}
The success of this attack is based on the inability of an entity to
get the latest updates on trust relations in the network, mainly due to
network attackers. Our blockchain abstraction provides a global view of
the trust graph and a partial ordering of trust relations with regard to time.
Therefore, all entities with access to the blockchain, i.e. to any single honest entity
in the network, are able to detect stale
information and discard it accordingly, thus making the stale information attack
impossible.
\subsection{Denial of Service attack}
\noindent\underline{The attack:}
The availability, i.e. usefulness, of a system is a major component of 
its ability to operate securely. \ac{DoS} attacks may 
not affect the core security properties of a TMS, in the sense 
that no illegal operation will be carried out, however, the operation 
of the real-world applications depending on the TMS is disrupted in 
various degrees depending on the application. There is recent evidence of
denial of service attacks on the internet's DNS infrastructure paralyzing a
large portion of the system \cite{dynattack}. The authentication infrastructure
is also a lucrative target for attackers. Certificate revocation services (X.509), as
well as keyservers (PGP) could be targeted in an attempt to hinder the functionality
of the system.
\smallskip
\noindent\\\underline{The defense:}
The decentralized nature of blockchain technology can make \ac{DoS} attacks
against specific entities ineffective. Specifically, under the assumptions of our
ideal blockchain model, any entity with network access to
the blockchain overlay is able to retrieve the latest, converged state of
the system and thus make informed trust assessments. The peers providing the
information need not be trusted, as security is guaranteed under the assumptions
outlined in Section \ref{sec:blockchain}. Network attacks against the blockchain
itself have been studied in \cite{vasek2014empirical} and \cite{heilman2015eclipse}
for the case of Bitcoin. These attacks generally require much stronger
adversaries than the ones considered in this paper, and are also accompanied
with mitigation measures.
\subsection{Censorship attack}
\noindent\underline{The attack:}
A censorship attack is essentially a \ac{DoS} variant. However, as
the objective and the motives differ from the general case of a \ac{DoS},
we handle this attack individually. In this scenario, a powerful adversary
forbids other entities within his legal jurisdiction to provide authentication
services to specific entities. Centralized systems (e.g. X.509) are inherently susceptible to this attack, however decentralized systems that rely on keyservers for trust
information availability (e.g. PGP) can also be affected.
\smallskip
\noindent\\\underline{The defense:}
Blockchain-based systems are assumed to be inherently censorship resistant, meaning that no single
entity, or small group thereof, can prevent other entities from submitting transactions
to the ledger. An adversary would need to control the majority of
participating nodes in order to mount this attack, which is
contrary to our security assumptions. However, when considering rational
adversaries, an entity may not want to risk producing a block that is in turn
rejected by the network, and therefore this attack may be more likely in this
setting. However, this adversary model is out of the scope of this paper.
\smallskip
\begin{remark}\label{rem:other}
	Different methods to enhance the security of the web PKI (X.509)
	system have been proposed in recent years \cite{galperin2013x,
		wendlandt2008perspectives, laurie2013certificate, melara2015coniks,
	syta2016keeping, braun2014trust, classen2015distributed}. Most of these approaches distribute trust among monitors
	and witnesses to keep the CAs honest. These enhancements are either
	inherently offered by blockchain constructions by design or can be
	implemented upon them.
\end{remark}
\subsection{Section summary}
In this section, we presented five prevalent attacks against \ac{TM} systems for
authentication. Our model (Section~\ref{sec:model}), allowed us to reason about
such systems in a generic manner, and in combination with our blockchain model,
enabled us to conceptually analyze the use of blockchain technology for \acl{TM}.
We showed that blockchain technology can mitigate, or completely negate the attacks
described in this section, therefore providing evidence of its merit in this area.

%% file: challenges.tex
\section{Discussion and challenges}\label{sec:challenges}
In this paper, we showcased the advantages of using \acp{ODL} in \ac{TM}
for authentication.
Despite ongoing research making first steps towards this direction,
there are still a multitude of problems to be resolved before
this technology can be widely deployed. In our blockchain abstraction
(Section \ref{sec:blockmodel}), we
implied that the whole set of state-changing transactions is stored
on the blockchain. The size of this construct will quickly cause the
blockchain to bloat \cite{wagner2014ensuring}, thus continuously increasing the capacity needs
of participating entities. Moreover, as the capacity and processing
needs are going to be substantial, thin clients \cite{frey2016bringing}
will need to be deployed on resource-constrained devices. When considering
the security of the blockchain itself, ardent theoretical and experimental
evaluation should accompany any proposed design. Regarding the last point,
the incentive mechanism used for blockchain participation will greatly influence the security of the
blockchain and the problem of designing it is of great significance. Privacy
is another major concern that has to be resolved, as balancing the
need for transparency and user privacy is a universal problem. Specifically,
the transparency that comes hand in hand with the use of blockchain technology
can provide information about the social relations and interactions of
parties, that needs to be protected. Another key design consideration is that
of the type of blockchain to be used. Public/open/permissionless designs, like the ones
adopted by Bitcoin and Ethereum, enable the open participation of all entities that 
want to contribute to the system. In this case, it will be important to hold the
participants accountable for their actions, so that they face repercussions if
they misbehave. On the other hand, so called permissioned/consortium blockchain
designs, as the one used in \cite{fabric}, can offer some advantages regarding
accountability, however they lack the decentralized nature of permissionless
systems.
\section{Conclusion}\label{sec:concl}
In this paper, we showed that \acp{ODL}, as implemented by blockchain technology, can
\textit{by design} enhance the security of authentication infrastructures.
We introduced an abstract, graph-theoretic model of \ac{TM} systems for
authentication, and a matching blockchain model. Our blockchain model
for \acl{TM} is a distributed, probabilistic state machine.
States of the machine correspond to snapshots of our graph-theoretic
model.
We then highlighted five
prevalent attacks and showed that under the assumptions of our
blockchain abstraction, they can be alleviated by encoding the
trust information in a secure blockchain. Specifically, the fact that blockchain
enables all participants to have a consistent, transparent view of the trust network,
solves many of the issues of traditional authentication systems.
By doing this, we showcased
that \textit{blockchain} is not just a buzzword in the case of \acl{TM} systems.
There is genuine merit in applying blockchain-based
designs in this field of network security, and that can lead to more
secure and trustworthy systems overall. The authentication problem is
expected to become even more challenging, as new paradigms, like the IoT,
come into reality and systematically exploring the advantages of blockchain
technology can provide new solutions. While blockchains and
\acp{ODL} in general, can offer concrete advantages compared with traditional approaches
to the problem of authentication, there are still important issues to be resolved
in order for this technology to yield real-world results.

%% file: acronyms.tex
\begin{acronym}
	\acro{id}{identity}
	\acro{ODL}{Open Distributed Ledger}
	\acro{MITM}{Man in the Middle}
	\acro{DoS}{Denial of Service}
	\acro{TM}{Trust Management}
\end{acronym}